\documentclass[12pt]{article}
\newcommand{\bq}{\begin{equation}}
\newcommand{\eq}{\end{equation}}
\newcommand{\ba}{\begin{eqnarray}}
\newcommand{\ea}{\end{eqnarray}}

\newcommand{\nl }{ \nonumber  }

\newcommand{\ul}{\underline}
\newcommand{\p}{\partial}
\newcommand{\pu}{\p_\tau}

\newcommand{\h}{\hspace{1cm}}
\newcommand{\s}{\sigma}
\newcommand{\us}{\underline\sigma}

\newcommand{\La}{\Lambda}
\newcommand{\pj}{\p_j}

\newcommand{\pk}{\p_k}

\def\be{\begin{eqnarray}}
\def\ee{\end{eqnarray}}

\begin{document}
\begin{center}
\vspace*{1.cm} {\bf EXACT BRANE SOLUTIONS IN CURVED BACKGROUNDS \footnote
{\it Talk at the Third International Conference on Geometry,
Integrability, and Quantization, June 14-23, 2001, Varna, Bulgaria. To be
published in the proceedings.}\\ \vspace*{.5cm} P. Bozhilov
\footnote{E-mails: p.bozhilov@shu-bg.net; bojilov@thsun1.jinr.ru}},
\\ {\it Department of Theoretical Physics,
\\ "Konstantin Preslavsky" University, 9712 Shoumen, Bulgaria}
\end{center}
\vspace*{.5cm}

We consider the classical null $p$-brane dynamics in $D$-dimensional
curved backgrounds and apply the Batalin-Fradkin-Vilkovisky approach for
BRST quantization of general gauge theories. Then we develop a method for
solving the tensionless $p$-brane equations of motion and constraints.
This is possible whenever there exists at least one Killing vector for the
background metric. It is shown that the same method can be also applied
for the {\it tensile} $1$-branes. Finally, we give two examples of
explicit exact solutions in four dimensions.


\vspace*{.5cm}

\section{Introduction}
\hspace{1cm} The $p$-brane is a $p$-dimensional relativistic object, which
evolving in space-time describes a $(p+1)$-dimensional worldvolume. In
this terminology, $p=0$ corresponds to a point particle, $p=1$ corresponds
to a string, $p=2$ corresponds to a membrane and so on. Every $p$-brane
characterizes by its tension $T_p$ with dimension of $(mass)^{p+1}$. When
the tension $T_p = 0$, the $p$-brane is called null or tensionless one.
This relationship between the null branes and the tensile ones generalizes
the correspondence between massless and massive particles for the case of
extended objects. Thus, the tensionless branes may be viewed as a
high-energy limit of the tensile ones.

As is known, there exist five consistent string theories in ten
dimensions: Type IIA with $N=2$ non-chiral supersymmetry, type IIB with
$N=2$ chiral supersymmetry, type I with N=1 supersymmetry and gauge
symmetry $SO(32)$ and heterotic strings with N=1 supersymmetry with
$SO(32)$ or $E_8\times E_8$ gauge symmetry.

The superstring dynamics unify all fundamental interactions between the
elementary particles, including gravity, at super high energies. The
$p$-branes arise naturally in the superstring theory, because there exist
exact brane solutions of the superstring effective equations of motion.
The $2$-branes and the $5$-branes are the fundamental dynamical objects in
eleven dimensional $M$-theory, which is the strong coupling limit of the
five superstring theories in ten dimensions, and which low energy field
theory limit is the eleven dimensional supergravity. Particular type of
$3$-branes arise in the Randall-Sundrum brane world scenario.

The purpose of this talk is to present some investigations on the
$p$-brane dynamics in curved backgrounds, which are part of the string
theory backgrounds, with the aim of finding exact solutions of the
equations of motion and constraints, and further application of the
received results. For example, if the branes are viewed as space-time
probes, the obtained exact solutions may have relevance to the singularity
structure of branes. On the other hand, these solutions may have
cosmological implications especially in the early universe. It is worth
checking if these solutions lead to self-consistent brane cosmology. The
possible application in the framework of the modern concept of {\it brane
world universe} is especially interesting. Another appropriate field of
realization of these results is the investigation of the solution
properties near black hole horizons, where the tensionless limit is a good
approximation and significantly simplifies the corresponding analysis. The
approach of Batalin, Fradkin and Vilkovisky for BRST quantization of
general gauge theories, applied to the null $p$-branes, gives the
possibility for quantization of such systems in curved backgrounds.

\section{\bf Null branes}
\subsection{\bf Lagrangian formulation}
\hspace{1cm}The action for the bosonic null $p$-brane in a D-dimensional
curved space-time with metric tensor $g_{MN}$ can be written in the form
\cite{PRD60}: \ba\label{a}S=\int d^{p+1}\xi {\cal L} ,\h {\cal
L}=V^mV^n\p_m X^M\p_n X^N g_{MN},\h\h
\\ \nl
\p_m=\p/\p\xi^m , \h \xi^m=(\xi^0,\xi^i)=(\tau,\s^i),\h\h
\\ \nl
m,n=0,1,...,p ,\h j,k=1,...,p ,\h M,N=0,1,...,D-1. \ea

To prove the invariance of the action under infinitesimal diffeomorphisms
on the world volume (reparametrizations), we first write down the
corresponding transformation law for the (r,s)-type tensor density of
weight $a$ \ba\nl \delta_{\varepsilon}T^{J_1...J_r}_{K_1...K_s}[a]&=&
L_{\varepsilon}T^{J_1...J_r}_{K_1...K_s}[a]= \varepsilon^L\p_L
T^{J_1...J_r}_{K_1...K_s}[a]\\ \label{diff} &+&
T^{J_1...J_r}_{KK_2...K_s}[a]\p_{K_1}\varepsilon^K+...+
T^{J_1...J_r}_{K_1...K_{s-1}K}[a]\p_{K_s}\varepsilon^K \\ \nl &-&
T^{JJ_2...J_r}_{K_1...K_s}[a]\p_J\varepsilon^{J_1}-...-
T^{J_1...J_{r-1}J}_{K_1...K_s}[a]\p_J\varepsilon^{J_r} \\ \nl &+&
aT^{J_1...J_r}_{K_1...K_s}[a]\p_L\varepsilon^L , \ea where $L_\varepsilon$
is the Lie derivative along the vector field $\varepsilon$. Using
(\ref{diff}), one verifies that if $X^M(\xi)$, $g_{MN}(\xi)$ are
world-volume scalars ($a=0$) and $V^m(\xi)$ is a world-volume (1,0)-type
tensor density of weight $a=1/2$, then $\p_m X^N$ is a (0,1)-type tensor,
$\p_m X^M \p_n X^N g_{M N}$ is a (0,2)-type tensor and ${\cal L}$ is a
scalar density of weight $a=1$. Therefore, \ba\nl
\delta_{\varepsilon}S=\int d^{p+1}\xi\p_m\bigl ( \varepsilon^m {\cal
L}\bigr ) \ea and this variation vanishes under suitable boundary
conditions.

The equations of motion following from (\ref{a}) are: \ba\nl \p_m\Bigl
(V^m V^n \p_n X^{L}\Bigr ) + \Gamma^{L}_{MN} V^m V^n \p_m X^{M}\p_n X^{N}
= 0 ,\\ \nl V^m \p_m X^{M}\p_n X^{N} g_{MN} = 0 , \ea where
$\Gamma^{L}_{MN}$ is the connection compatible with the metric $g_{MN}$:
\ba\nl \Gamma^{L}_{MN}=\frac{1}{2}g^{LR}\bigl(
\p_{M}g_{NR}+\p_{N}g_{MR}-\p_{R}g_{MN}\bigr) . \ea

For the transition to Hamiltonian picture it is convenient to rewrite the
Lagrangian density (\ref{a}) in the form ($\pu=\p/\p\tau, \pj=\p/\p\s^j$):
\ba\label{L} L=\frac{1}{4\lambda^0} g_{MN}\bigl (\pu-\lambda^j\pj\bigr
)X^M \bigl (\pu-\lambda^k\pk\bigr )X^N , \ea where \ba\nl
V^m=\bigl(V^0,V^j\bigr)=\Biggl(-\frac{1}{2\sqrt{\lambda^0}},
\frac{\lambda^j}{2\sqrt{\lambda^0}}\Biggr) . \ea Now the equation of
motion for $X^N$ takes the form: \ba\nl
\pu\Bigl
[\frac{1}{2\lambda^0}\bigl (\pu-\lambda^k\pk\bigr )X^{L}\Bigr ] -\pj\Bigl
[\frac{\lambda^{j}}{2\lambda^{0}}\bigl (\pu-\lambda^k\pk\bigr )X^{L} \Bigr
]
\\ \nl + \frac{1}{2\lambda^0}\Gamma^{L}_{MN} \bigl(\pu-\lambda^j
\pj\bigr)X^M \bigl (\pu-\lambda^k\pk\bigr )X^N = 0 . \ea The equations of
motion for the Lagrange multipliers $\lambda^{0}$ and $\lambda^{j}$ which
follow from (\ref{L}) give the constraints : \ba\nl
g_{MN}\bigl
(\pu-\lambda^j\pj\bigr )X^M \bigl (\pu-\lambda^k\pk\bigr )X^N = 0 , \\ \nl
g_{MN}\bigl (\pu-\lambda^k\pk\bigr )X^M \pj X^N = 0 . \ea In terms of
$X^N$ and the conjugated momentum $P_N$ they read: \ba\label{Tpx}
T_0=g^{MN}P_M P_N = 0 \h,\h T_j=P_N\pj X^N = 0 . \ea

\subsection{\bf Hamiltonian formulation}
\hspace{1cm} The Hamiltonian which corresponds to the Lagrangian density
(\ref{L}) is a linear combination of the constraints (\ref{Tpx}) : \ba\nl
H_0=\int d^p\sigma\bigl (\lambda^0 T_0+\lambda^j T_j \bigr ) . \ea They
satisfy the following (equal $\tau$) Poisson bracket algebra \ba \nl
\{T_0(\ul \sigma_1),T_0(\ul \sigma_2)\}&=&0,
\\ \label {CA}
\{T_0(\ul \sigma_1),T_{j}(\ul \sigma_2)\}&=& [T_0(\ul \sigma_1) + T_0(\ul
\sigma_2)] \p_j \delta^p (\ul \sigma_1 - \ul \sigma_2) ,
\\ \nl
\{T_{j}(\ul \sigma_1),T_{k}(\ul \sigma_2)\}&=& [\delta_{j}^{l}T_{k}(\ul
\sigma_1) + \delta_{k}^{l}T_{j}(\ul \sigma_2)]\p_l\delta^p(\ul
\sigma_1-\ul \sigma_2) ,\\ \nl \us=(\s^1,...,\s^p) . \ea The equalities
(\ref{CA}) show that the constraint algebra is the same for flat and for
curved backgrounds. Having in mind the above algebra, one can use the
Batalin-Fradkin-Vilkovisky approach for BRST quantization of general gauge
theories, and to construct the corresponding BRST charge $\Omega$
(*=complex conjugation) \ba \label{O} \Omega = \Omega^{min}+\pi_m \bar
{\cal P}^m , \h \{\Omega,\Omega\} = 0 , \h \Omega^* = \Omega . \ea
$\Omega^{min}$ in (\ref{O}) can be written as \cite{PRD60}  \ba\nl
\Omega^{min}&=&\int d^p\s\{T_0\eta^0+T_j\eta^j+ {\cal P}_0
[(\p_j\eta^j)\eta^0 + (\p_j\eta^0)\eta^j ] + {\cal P}_k(\p_j\eta^k)\eta^j
\} , \ea and can be represented also in the form \ba\nl \Omega^{min}=\int
d^p\s\bigl [\bigl (T_0+ \frac{1}{2}T_0^{gh}\bigr )\eta^0 +\bigl
(T_j+\frac{1}{2}T_j^{gh}\bigr )\eta^j \bigr ] +\int d^p\s\p_j\Bigl
(\frac{1}{2}{\cal P}_k\eta^k\eta^j \Bigr ) . \ea Here a superscript $gh$
is used for the ghost part of the total gauge generators \ba\nl \nl
T_m^{tot}=\{\Omega,{\cal P}_m\}=\{\Omega^{min},{\cal P}_m\}= T_m+T_m^{gh}
. \ea We recall that the Poisson bracket algebras of $T_m^{tot}$ and $T_m$
coincide for first rank systems which is the case under consideration. The
manifest expressions for $T_m^{gh}$ are: \ba\nl T_0^{gh}&=&2{\cal
P}_0\p_j\eta^j+\bigl (\p_j{\cal P}_0\bigr )\eta^j ,
\\ \nl
T_j^{gh}&=&2{\cal P}_0\p_j\eta^0+\bigl (\p_j{\cal P}_0\bigr )\eta^0+ {\cal
P}_j\p_k\eta^k+{\cal P}_k\p_j\eta^k+ \bigl (\p_k{\cal P}_j\bigr )\eta^k .
\ea Up to now, we introduced canonically conjugated ghosts $\bigl
(\eta^m,{\cal P}_m\bigr )$, $\bigl (\bar \eta_m,\bar {\cal P}^m\bigr)$ and
momenta $\pi_m$ for the Lagrange multipliers $\lambda^m$ in the
Hamiltonian. They have Poisson brackets and Grassmann parity as follows
($\epsilon_m$ is the Grassmann parity of the corresponding constraint):
\ba\nl \bigl \{\eta^m,{\cal P}_n\bigr \}&=&\delta^m_n , \h \epsilon
(\eta^m)=\epsilon ({\cal P}_m)=\epsilon_m + 1 , \\ \nl \bigl \{\bar
\eta_m,\bar {\cal P}^n \bigr \}&=&-(-1)^{\epsilon_m\epsilon_n} \delta^n_m
, \h \epsilon (\bar\eta_m)=\epsilon (\bar {\cal P}^m)=\epsilon_m + 1 ,
\\ \nl
\bigl \{\lambda^m,\pi_n\bigr \}&=&\delta^m_n , \h \epsilon
(\lambda^m)=\epsilon (\pi_m)=\epsilon_m . \ea

The BRST-invariant Hamiltonian is \ba\label{H} H_{\tilde
\chi}=H^{min}+\bigl \{\tilde \chi,\Omega\bigr \}= \bigl \{\tilde
\chi,\Omega\bigr \} , \ea because from $H_{canonical}=0$ it follows
$H^{min}=0$. In this formula $\tilde \chi$ stands for the gauge fixing
fermion $(\tilde \chi^* = -\tilde \chi)$. We use the following
representation for the latter \ba\nl \tilde
\chi=\chi^{min}+\bar\eta_m(\chi^m+\frac{1}{2}\rho_{(m)}\pi^m) , \h
\chi^{min}=\lambda^m{\cal P}_m, \ea where $\rho_{(m)}$ are scalar
parameters and we have separated the $\pi^m$-dependence from $\chi^m$. If
we adopt that $\chi^m$ does not depend on the ghosts $(\eta^m,{\cal P}_m)$
and $(\bar\eta_m,\bar {\cal P}^m)$, the Hamiltonian $H_{\tilde\chi}$ from
(\ref{H}) takes the form \ba \label{r}
H_{\tilde\chi}&=&H_{\chi}^{min}+{\cal P}_m \bar {\cal P}^m -
\pi_m(\chi^m+\frac{1}{2}\rho_{(m)}\pi^m)+ \\ \nl &+&\bar\eta_m \bigl
\{\chi^m,T_n\bigr \}\eta^n , \ea where \ba\nl H_{\chi}^{min}=\bigl
\{\chi^{min},\Omega^{min}\bigr \} . \ea

One can use the representation (\ref{r}) for $H_{\tilde\chi}$ to obtain
the corresponding BRST invariant Lagrangian density \ba\nl
L_{\tilde\chi}=L+L_{GH}+L_{GF} . \ea Here $L$ is given in (\ref{L}),
$L_{GH}$ stands for the ghost part and $L_{GF}$ - for the gauge fixing
part of the Lagrangian density. The manifest expressions for $L_{GH}$ and
$L_{GF}$ are \cite{PRD60}: \ba \nl
L_{GH}=-\p_\tau\bar{\eta_0}\p_\tau\eta^0-\p_\tau\bar{\eta_j}
\p_\tau\eta^j+\lambda^0[2\pu\bar{\eta_0}\pj\eta^j+
(\pj\pu\bar{\eta_0})\eta^j]
\\
\nl +\lambda^j[2\pu\bar{\eta_0}\pj\eta^0+
(\pj\pu\bar{\eta_0})\eta^0+\pu\bar{\eta_k}\pj\eta^k+
\pu\bar{\eta_j}\p_k\eta^k+ (\p_k\pu\bar{\eta_j})\eta^k]
\\
\nl +\int d^p\sigma'\{\bar{\eta_0}(\underline{\sigma'})
[\{T_0,\chi^0(\underline{\sigma'})\}\eta^0
+\{T_j,\chi^0(\underline{\sigma'})\}\eta^j]
\\
\nl +\bar{\eta_j}(\underline{\sigma'})[\{T_0,\chi^j(\underline{\sigma'})\}
\eta^0+\{T_k,\chi^j(\underline{\sigma'})\}\eta^k]\} , \\ \nl
L_{GF}=\frac{1}{2\rho_{(0)}}(\p_\tau \lambda^0-\chi^0)(\p_\tau
\lambda_0-\chi_0)+ \frac{1}{2\rho_{(j)}}(\p_\tau \lambda^j-\chi^j)
(\p_\tau \lambda_j-\chi_j) . \ea

If one does not intend to pass to the Lagrangian formalism, one may
restrict oneself to the minimal sector $\bigl
(\Omega^{min},\chi^{min},H_\chi^{min}\bigr )$. In particular, this means
that Lagrange multipliers are not considered as dynamical variables
anymore. With this particular gauge choice, $H_\chi^{min}$ is a linear
combination of the total constraints \ba\nl H_\chi^{min}=\int d^p\s\Bigl
[\La^0 T_0^{tot}(\us)+\La^j T_j^{tot}(\us)\Bigr ] , \ea and we can treat
here the Lagrange multipliers $\La^0,\La^{j}$ as constants.

As a result, we have the possibility to quantize this dynamical system
living in curved background.

\subsection{Solving the equations of motion}
\hspace{1cm}The brane equations of motion and constraints in curved
space-time are highly nonlinear and, {\it in general}, non exactly
solvable. Different methods have been applied to solve them approximately
or, if possible, exactly in a fixed background. On the other hand, quite
general exact solutions can be found by using an appropriate ansatz, which
exploits the symmetries of the underlying curved space-time
\cite{PRD60,PLB472,PRD62}. We will use namely this approach.

From now on, we will work in the gauge $\lambda^m = constants$, in which
the equations of motion may be written in the form \ba\nl
g_{LN}\bigl(\pu-\lambda^j\p_j\bigr)^2 X^{N} + \Gamma_{L,MN} \bigl
(\pu-\lambda^j\p_j\bigr )X^M \bigl (\pu-\lambda^k\p_k\bigr )X^N = 0 . \ea
First of all, we will look for background independent solution of these
equations and of the constraints
 \ba\label{tc1} g_{MN}\bigl
(\pu-\lambda^j\pj\bigr )X^M \bigl (\pu-\lambda^k\pk\bigr )X^N = 0 , \\
\label{tc2} g_{MN}\bigl (\pu-\lambda^k\pk\bigr )X^M \pj X^N = 0 . \ea It
is easy to check that the solution is \cite{PRD60,PLB472} \ba\nl X^M
(\xi)= F^M(\lambda^i\xi^0 + \xi^i) \equiv  F^M(\lambda^i\tau +
\sigma^i),\ea where $F^M$ are $D$ arbitrary functions of their arguments.
The next step is to use the existing symmetries of the background metric.
To this end, let us split the index $M=(\mu,a)$,
$\{\mu\}\neq\{\emptyset\}$ and let us suppose that there exist a number of
independent Killing vectors $\eta_{\mu}$. Then in appropriate coordinates
$\eta_{\mu}=\frac{\p}{\p x^{\mu}}$ and the metric does not depend on
$X^{\mu}$. Now our aim is to find an ansatz for $X^M$, which will allow us
to separate the variables $\xi^0$ and $\xi^i$ on the one hand, and on the
other hand - to find the first integrals of a part of the equations of
motion, corresponding to the symmetry of the curved background space-time.
It turns out that the appropriate ansatz is: \ba\nl
X^{\mu}\left(\tau,\sigma^i\right) &=& C^{\mu}F(\lambda^i\tau + \sigma^i) +
y^{\mu}\left(\tau\right),\h C^{\mu}=constants,
\\ \nl  X^{a}\left(\tau,\sigma^i\right) &=& y^{a}\left(\tau\right).\ea
Inserting it in the equations of motion and constraints, one obtains the
conserved quantities : \ba\nl
g_{\mu\nu}\dot{y}^{\nu} + g_{\mu
a}\dot{y}^a = A_{\mu} = constants.\ea The constraints (\ref{tc2}) are
identically satisfied when $A_{\mu} C^{\mu} = 0$. The remaining equations
and the constraint (\ref{tc1}) are reduced to \ba\nl g_{aN}\ddot{y}^N +
\Gamma_{a,MN}\dot{y}^M\dot{y}^N = 0\\ \nl
 g_{MN}\dot{y}^M\dot{y}^N = 0.\ea

Using the obtained first integrals, these equalities can be rewritten as
\ba\nl
2\frac{d}{d\tau}\left(h_{ab}\dot{y}^b\right) - \left(\p_a
h_{bc}\right)\dot{y}^b\dot{y}^c + \p_a V = 4\p_{[a}A_{b]}\dot{y}^b\\
\label{c2a} h_{ab}\dot{y}^a\dot{y}^b + V =0, \ea where \ba\nl h_{ab}\equiv
g_{ab} - g_{a\mu}k^{\mu\nu}g_{\nu b},\h V\equiv A_{\mu}A_{\nu}k^{\mu\nu}
\h A_{a} \equiv g_{a\mu}k^{\mu\nu}A_{\nu} ,\ea and $k^{\mu\nu}$ is by
definition the inverse of $g_{\mu\nu}$:
$k^{\mu\lambda}g_{\lambda\nu}=\delta^{\mu}_{\nu}$. Thus, using the
existence of an abelian isometry group $G$ generated by the Killing
vectors $\p/\p x^{\mu}$, the problem of solving the equations of motion
and $p+1$ constraints in $D$-dimensional curved space-time $\mathcal{M}_D$
with metric $g_{MN}$ is reduced to considering equations of motion and one
constraint in the coset $\mathcal{M}_D/G$ with metric $h_{ab}$. As might
be expected, an interaction with an effective gauge field appears in the
Euler-Lagrange equations. In this connection, let us note that if we write
down $A_{a}$ as \ba\nl A_{a}=A_a^{\nu}A_{\nu},\ea this establishes a
correspondence with the usual Kaluza-Klein type notations and \ba\nl
g_{MN}dy^M dy^N = h_{ab}dy^a dy^b + g_{\mu\nu} \left(dy^\mu + A^\mu_a
dy^a\right)\left(dy^\nu + A^\nu_b dy^b\right).\ea

At this stage, we restrict the metric $h_{ab}$ to be a diagonal one, i.e.
\ba\label{mc} g_{ab} = g_{a\mu}k^{\mu\nu}g_{\nu b}, \h\h \mbox{for}\h\h
a\ne b.\ea This allows us to transform further the equations for $y^a$ and
obtain \ba\label{emat} && \frac{d}{d\tau}\left(h_{aa}\dot{y}^a\right)^2 +
\dot{y}^a\p_a\left(h_{aa} V\right) \\ \nl &&+ \dot{y}^a\sum_{b\ne a}
\left[\p_a\left(\frac{h_{aa}}{h_{bb}}\right)\left(h_{bb}\dot{y}^b\right)^2
- 4\p_{[a}A_{b]} h_{aa}\dot{y}^b\right] = 0.\ea

To reduce the order of the differential equations (\ref{emat}) by one, we
first split the index $a$ in such a way that $y^r$ is one of the
coordinates $y^a$, and $y^{\alpha}$ are the others. Then we impose the
sufficient conditions \ba\label{cond}
\p_{\alpha}\left(\frac{h_{\alpha\alpha}}{h_{aa}}\right)=0,\h
\p_{\alpha}\left(h_{rr}\dot{y}^r\right)^2 = 0,\\ \nl
\p_{r}\left(h_{\alpha\alpha}\dot{y}^{\alpha}\right)^2 = 0,\h
A_{\alpha}=\p_{\alpha}f.\ea The result of integrations, compatible with
(\ref{c2a}) and (\ref{mc}), is the following \ba\nl
\left(h_{\alpha\alpha}\dot{y}^{\alpha}\right)^2 = D_{\alpha} \left(y^a\ne
y^{\alpha}\right) + h_{\alpha\alpha}\left[2\left( A_{r}-\p_r
f\right)\dot{y}^r - V\right]=E_{\alpha} \left(y^{\beta}\right),\\
\label{fia} \left(h_{rr}\dot{z}^{r}\right)^2 =
h_{rr}\left\{\left(\sum_{\alpha}-1\right) V -
\sum_{\alpha}\frac{D_{\alpha}}{h_{\alpha\alpha}}\right\} +
\left[\sum_{\alpha}\left(A_{r}-\p_r f\right)\right]^2 =
E_r\left(y^r\right),\ea where $D_{\alpha}$, $E_{\alpha}$, $E_r$ are
arbitrary functions of their arguments, and \ba\nl \dot{z}^r \equiv
\dot{y}^r + \frac{\sum_{\alpha}}{h_{rr}} \left(A_{r}-\p_r f\right).\ea To
find solutions of the above equations without choosing particular metric,
we have to fix all coordinates $y^a$ except one. If we denote it by $y^A$,
then the $exact$ solutions of the equations of motion and constraints for
a null $p$-brane in the considered curved background are given by \ba\nl
&&X^{\mu}\left(X^A,\s^j\right)=X^{\mu}_{0} +
C^{\mu}F\left(\lambda^j\tau+\sigma^j\right) \\ \nl &&-
\int_{X_0^A}^{X^A}k_0^{\mu\nu}\left[g^0_{\nu A}\mp A_{\nu}
\left(-\frac{h^0_{AA}}{V^{0}}\right)^{1/2}\right]d u ,\\ \nl
&&\tau\left(X^A\right)=\tau_0 \pm \int_{X_0^A}^{X^A}
\left(-\frac{h^0_{AA}}{V^{0}}\right)^{1/2}d u.\ea

\section{\bf Exact solutions for the tensile 1-brane}
\hspace{1cm}To begin with, we write down the bosonic string action in $D$-
dimensional curved space-time $\mathcal{M}_D$ with metric tensor $g_{MN}$
\ba\nl S&=&\int d^{2}\xi {\cal L},\h {\cal L}=
-\frac{T}{2}\sqrt{-\gamma}\gamma^{mn} \p_m X^M\p_n X^N
g_{MN}\left(X\right), \ea where, as usual, $T$ is the string tension and
$\gamma$ is the determinant of the auxiliary metric $\gamma_{mn}$.

Here we would like to consider tensile and null (tensionless) strings on
equal footing, so we have to rewrite the action  in a form in which the
limit $T\to 0$ could be taken. To this end, we set \cite{0103154} \ba\nl
\gamma^{mn}= \left(\begin{array}{cc}-1&\lambda^1\\
\lambda^1&-\left(\lambda^1\right)^2 + \left(2\lambda^0
T\right)^2\end{array}\right)\ea and obtain \ba\nl &&{\cal L}=
\frac{1}{4\lambda^0}g_{MN}\left(X\right) \left(\p_0-\lambda^1\p_1\right)
X^M\left(\p_0-\lambda^1\p_1\right) X^N \\ \nl && - \lambda^0 T^2
g_{MN}\left(X\right)\p_1X^M\p_1X^N.\ea The equations of motion and
constraints following from this Lagrangian density are ($\lambda^m =
constants$):\ba\nl && g_{LN}\left(\p_0-\lambda^1\p_1\right)^2 X^N +
\Gamma_{L,MN} \left(\p_0-\lambda^1\p_1\right) X^M
\left(\p_0-\lambda^1\p_1\right) X^N \\ \nl &&= \left(2\lambda^0 T\right)^2
\left( \p_1^2X^K + \Gamma^{K}_{MN}\p_1X^M\p_1X^N \right),\\ \nl
&&g_{MN}\left(X\right) \left(\p_0-\lambda^1\p_1\right) X^M
\left(\p_0-\lambda^1\p_1\right) X^N \\ \nl &&+ \left(2\lambda^0 T\right)^2
g_{MN}\left(X\right)\p_1X^M\p_1X^N = 0,\\ \nl
&&g_{MN}\left(X\right)\left(\p_0-\lambda^1\p_1\right) X^M\p_1X^N=0. \ea

The background independent solution of the equations of motion (but not of
the constraints) is \cite{0103154}\ba\nl X^M(\tau,\sigma) =
F^{M}_{\pm}[w_{\pm}(\tau,\sigma)],\h
w_{\pm}(\tau,\sigma)=(\lambda^1\pm2\lambda^0 T)\tau + \sigma. \ea The
ansatz with the searched properties is given by \ba\nl
X^{\mu}\left(\tau,\sigma\right) &=& C^{\mu}_{\pm}w_{\pm} +
y^{\mu}\left(\tau\right),\h C^{\mu}_{\pm}=constants,
\\ \nl  X^{a}\left(\tau,\sigma\right) &=& y^{a}\left(\tau\right).\ea
Applying this ansatz for the equations of motion and constraints one
obtains \ba\nl g_{KL}\ddot{y}^L + \Gamma_{K,MN}\dot{y}^M\dot{y}^N \pm
4\lambda^0 TC^{\mu}_{\pm}\Gamma_{K,\mu N}\dot{y}^N = 0.\ea \ba\nl
g_{MN}(y^a)\dot{y}^M \dot{y}^N \pm 2\lambda^0 TC^{\mu}_{\pm} \left[g_{\mu
N}(y^a)\dot{y}^N \pm 2\lambda^0 TC^{\nu}_{\pm} g_{\mu\nu}(y^a)\right] =
0,\\ \nl C^{\mu}_{\pm} \left[g_{\mu N}(y^a)\dot{y}^N \pm 2\lambda^0
TC^{\nu}_{\pm} g_{\mu\nu}(y^a)\right] = 0.\ea Obviously, this system of
two constraints is equivalent to the following one \ba\nl
g_{MN}(y^a)\dot{y}^M \dot{y}^N = 0,\\ \nl C^{\mu}_{\pm} \left[g_{\mu
N}(y^a)\dot{y}^N \pm 2\lambda^0 TC^{\nu}_{\pm} g_{\mu\nu}(y^a)\right] =
0,\ea which we will use from now on.

The integration of a part of the Euler-Lagrange equations leads to the
conserved quantities \cite{0103154}  \ba\nl g_{\mu\nu}\dot{y}^{\nu} +
g_{\mu a}\dot{y}^a \pm 2\lambda^0 TC^{\nu}_{\pm}g_{\mu\nu} = A^{\pm}_{\mu}
= constants,\h A^{\pm}_{\mu} C^{\mu}_{\pm} = 0.\ea Using the obtained
first integrals, one reduces the problem to solving the equations \ba\nl
2\frac{d}{d\tau}\left(h_{ab}\dot{y}^b\right) - \left(\p_a
h_{bc}\right)\dot{y}^b\dot{y}^c + \p_a V^\pm =
4\p_{[a}A^\pm_{b]}\dot{y}^b\\ \nl h_{ab}\dot{y}^a\dot{y}^b + V^\pm =0, \ea
where \ba\nl V^{\pm}\equiv A^{\pm}_{\mu}A^{\pm}_{\nu}k^{\mu\nu} +
\left(2\lambda^0 T\right)^2C^{\mu}_{\pm}C^{\nu}_{\pm}g_{\mu\nu},\h
A^{\pm}_{a} \equiv g_{a\mu}k^{\mu\nu}A^{\pm}_{\nu}.\ea Obviously, these
equalities have the same form as before with the replacements $V \to
V^\pm$, $A_a \to A_a^\pm$. Therefore, we can write down the exact solution
in this case right now, and it is (under the same conditions on the
metric) :\ba\nl &&X^{\mu}\left(X^A,\s\right)=X^{\mu}_{0} +
C^{\mu}_{\pm}\left(\lambda^1\tau+\sigma\right)\\ \nl &&-
\int_{X_0^A}^{X^A}k_0^{\mu\nu}\left[g^0_{\nu A}\mp A^{\pm}_{\nu}
\left(-\frac{h^0_{AA}}{V^{\pm 0}}\right)^{1/2}\right]d u ,\\ \nl &&
\tau\left(X^A\right)=\tau_0 \pm \int_{X_0^A}^{X^A}
\left(-\frac{h^0_{AA}}{V^{\pm 0}}\right)^{1/2}d u.\ea

\section{\bf Explicit examples}
\hspace{1cm} First of all, let us write down the generic structure of the
solutions as functions of the coordinate $X^A$. For tensionless
$p$-branes, i.e. $T=0$, $p$ - arbitrary, it is \ba\nl
&&X^{\mu}\left(X^A,\s^j\right)=X^{\mu}_{0} +
C^{\mu}F\left(\lambda^j\tau+\sigma^j\right)\pm\lim_{T\to 0}
I^{\mu}_{\pm}\left(X^A;T\right)\\ \nl &&\tau\left(X^A\right)=\tau_0
\pm\lim_{T\to 0} I_{0}^{\pm}\left(X^A;T\right).\ea For tensile strings,
i.e. $T\ne 0$, $p=1$, we have  \ba\nl
&&X^{\mu}\left(X^A,\s\right)=X^{\mu}_{0} +
C^{\mu}_{\pm}\left(\lambda^1\tau+\sigma\right) \pm
I^{\mu}_{\pm}\left(X^A;T\right)\\ \nl &&\tau\left(X^A\right)=\tau_0 \pm
I_{0}^{\pm}\left(X^A;T\right).\ea In our examples below, we will give the
expressions for $I^{\mu}_{\pm}$ and $I_{0}^{\pm}$.

Let us first consider an exact null $p$-brane solution for a four
dimensional cosmological {\it Kasner background}. Namely, the line element
is ($x^0 \equiv t$) \ba\nl ds^2 &=& g_{MN}dx^M dx^N = -(d t)^2 +
\sum_{\mu=1}^{3}t^{2q_{\mu}}(dx^{\mu})^2,\\ \label{Kc}
&&\sum_{\mu=1}^{3}q_{\mu}=1,\h \sum_{\mu=1}^{3}q^2_{\mu}=1.\ea

Without using the Kasner constraints (\ref{Kc}), the solution is given by
\cite{0103154} \ba\nl &&I^\mu(t)= constant
-\frac{\sqrt{\pi}}{A_2^{\pm}}\sum_{k=0}^{\infty}
\frac{\left(A_3^{\pm}/A_2^{\pm}\right)^{2k}}{k!\Gamma\left(1/2-k\right)}
\frac{t^{\mathcal{P}}}{\mathcal{P}}\times \\ \nl
&&\mbox{}_2F_1\left(1/2+k,\frac{ \mathcal{P}}{2(q_2-q_1)};
\frac{2(q_2-q_3)k+3q_2-2q_1+1-2q_\mu}{2(q_2-q_1)};
-\left(\frac{A_1^{\pm}}{A_2^{\pm}}\right)^2 t^{2(q_2-q_1)}\right),\\ \nl
&& \mathcal{P}\equiv 2(q_2-q_3)k+q_2+1-2q_\mu ,\h q_\mu = (q_1,q_2,q_3),\h
\mbox{for $q_1 > q_2$,} \ea  and \ba\nl &&I^\mu(t)= constant
+\frac{\sqrt{\pi}}{A_1^{\pm}}\sum_{k=0}^{\infty}
\frac{\left(A_3^{\pm}/A_1^{\pm}\right)^{2k}}{k!\Gamma\left(1/2-k\right)}
\frac{t^{\mathcal{Q}}}{\mathcal{Q}}\times \\ \nl
&&\mbox{}_2F_1\left(1/2+k,\frac{ \mathcal{Q}}{2(q_1-q_2)};
\frac{2(q_1-q_3)k+3q_1-2q_2+1-2q_\mu}{2(q_1-q_2)};
-\left(\frac{A_2^{\pm}}{A_1^{\pm}}\right)^2 t^{2(q_1-q_2)}\right),\\ \nl
&& \mathcal{Q}\equiv 2(q_1-q_3)k+q_1+1-2q_\mu,\h\mbox{for $q_1 < q_2$},\ea
where $\mbox{}_2F_1\left(a,b;c;z\right)$ is the Gauss' hypergeometric
function and $\Gamma(z)$ is the Euler's $\Gamma$-function. The expressions
for $I_{0}^{\pm}$ are obtainable from the above ones by setting $q_\mu =
0$. Because there are no restrictions on $q_\mu$, except $q_1\neq q_2\neq
q_3$, this probe brane solution is also valid in generalized Kasner type
backgrounds arising in superstring cosmology \cite{LWC99}. In string
frame, the effective Kasner constraints for the four dimensional
dilaton-moduli-vacuum solution are \ba\nl
&&\sum_{\mu=1}^{3}q_{\mu}=1+\mathcal{K},\h
\sum_{\mu=1}^{3}q^2_{\mu}=1-\mathcal{B}^2,\\ \nl
&&-1-\sqrt{3\left(1-\mathcal{B}^2\right)}\le \mathcal{K}\le
-1+\sqrt{3\left(1-\mathcal{B}^2\right)},\h
\mathcal{B}^2\in\left[0,1\right].\ea In Einstein frame, the metric has the
same form, but in new, rescaled coordinates and with new powers
$\tilde{q}_{\mu}$ of the scale factors. The generalized Kasner constraints
are also modified as follows \ba\nl &&\sum_{\mu=1}^{3}\tilde{q}_{\mu}=1,\h
\sum_{\mu=1}^{3}\tilde{q}^2_{\mu}=1-\tilde{\mathcal{B}}^2 -
\frac{1}{2}\tilde{\mathcal{K}}^2,\h \tilde{\mathcal{B}}^2 +
\frac{1}{2}\tilde{\mathcal{K}}^2\in \left[0,1\right] .\ea Actually, the
obtained tensionless $p$-brane solution is also relevant to considerations
within a {\it pre-big bang} context, because there exist a class of models
for pre-big bang cosmology, which is a particular case of the given
generalized Kasner backgrounds \cite{LWC99}.

Our next example is for a tensile string, evolving in the {\it Kerr
space-time} with metric taken in the form \ba\nl
g_{00}&=&-\Biggl(1-\frac{2Mr}{\rho^2}\Biggr),\h
g_{11}=\frac{\rho^2}{\Delta},\h g_{22}=\rho^2, \\ \nl
g_{33}&=&\Biggl(r^2+a^2+\frac{2Ma^2
r\sin^2\theta}{\rho^2}\Biggr)\sin^2\theta ,\h
g_{03}=-\frac{2Mar\sin^2\theta}{\rho^2} , \ea where \ba\nl
\rho^2=r^2+a^2\cos^2\theta \h,\h \Delta=r^2-2Mr+a^2 , \ea $M$ is the mass
and $a$ is the angular momentum per unit mass of the Kerr black hole. Now
the metric does not depend on $x^0$ and $x^3$, so that $\mu=0,3$, $a=1,2$.
The corresponding exact string solution, as a function of the radial
coordinate $r$, is described by the integrals
$\left(\theta=\theta_0=constant\right)$: \ba\nl
&&I^{0}_{\pm}\left(r;T\right)= \int_{r_0}^{r}dr \left[A_0^\pm
a^2\Delta\sin^2\theta_0-2A_3^\pm Mar-A_0^\pm\left(r^2+a^2\right)^2\right]
\left[-\Delta^3D_2\left(r;T\right)\right]^{-1/2},\\ \nl
&&I^{3}_{\pm}\left(r;T\right)= \int_{r_0}^{r}dr \left(
\frac{A_3^\pm\Delta}{\sin^2\theta_0}-2A_0^\pm Mar-A_3^\pm a^2\right)
\left[-\Delta^3D_2\left(r;T\right)\right]^{-1/2},\\ \nl
&&I_{0}^{\pm}\left(r;T\right)= \int_{r_0}^{r}dr \rho^2_0 \left[-\Delta
D_2\left(r;T\right)\right]^{-1/2},\h \rho^2_0=r^2+a^2\cos^2\theta_0,\\ \nl
&&D_2\left(r;T\right)=
-\frac{1}{\Delta}\left[\left(A_0^\pm\right)^2\left(r^2+a^2\right)^2 +
4A_0^\pm A_3^\pm Mar+\left(A_3^\pm\right)^2 a^2\right]\\ \nl
&&+\left(A_0^\pm\right)^2 a^2\sin^2\theta_0
+\frac{\left(A_3^\pm\right)^2}{\sin^2\theta_0} +\left(2\lambda^0T\right)^2
\left(C^0_\pm\right)^2\left(a^2\sin^2\theta_0-\Delta\right)\\ \nl
&&+\left(2\lambda^0T\right)^2\sin^2\theta_0
\left[\left(C^3_\pm\right)^2\left(r^2+a^2\right)^2 -4C^0_\pm C^3_\pm Mar
-\left(C^3_\pm\right)^2 a^2 \Delta\sin^2\theta_0 \right].\ea Analogously,
one can write down the solution as a function of $\theta$ when the radial
coordinate is kept fixed. Moreover, in the zero tension limit, one can
find the orbit $r=r(\theta)$, which is given by: \ba\nl \int_{r_0}^{r}
\frac{dr}{\Delta(r)E_1^{1/2}(r)}=\pm\int_{\theta_0}^{\theta}
\frac{d\theta}{E_2^{1/2}(\theta)},\ea where \ba\nl
&&E_1(r)=\frac{1}{\Delta^2}\left[A_0^2 \left(r^2+a^2\right)^2+4A_0 A_3
Mar+A_3^2 a^2 \right]-\frac{d}{\Delta},\\ \nl &&E_2(\theta)=d-\left(A_0^2
a^2\sin^2\theta+\frac{A_3^2}{\sin^2\theta}\right),\h d=constant.\ea The
possibility to obtain this result is due to the fact that on the one hand
the conditions (\ref{cond}) are satisfied, and on the other hand that in
(\ref{fia}) the variables $r$ and $\theta$ can be separated. This
corresponds to the separation of the variables in the Hamilton-Jakobi
equation for the Kerr metric, connected to the existence of a nontrivial
Killing tensor of second rank for this space-time.

\section{\bf Concluding remarks}
\hspace{1cm} In this talk we gave a short review of the results on the
$p$-brane dynamics in curved backgrounds, received in
\cite{PRD60}-\cite{0103154}. The new point here is the generalization of
the previously obtained null $p$-brane solutions to the case of more
general class of background metrics. The tensile $1$-brane exact solution
in Kerr space-time is also new one.

\vspace*{.5cm}
{\bf Acknowledgment}

This work was supported in part by a Shoumen University grant under
contract $No.20/2001$.


\end{document}